# Web Document Clustering and Ranking using Tf-Idf based Apriori Approach

Rajendra Kumar Roul
BITS, Pilani
K.K.Birla Goa Campus
Zuarinagar, Goa-403726, India
rkroul@goa.bits-pilani.ac.in

Omanwar Rohit Devanand
BITS, Pilani
K.K.Birla Goa Campus
Zuarinagar, Goa-403726, India
rohit.omanwar@gmail.com

S. K. Sahay
BITS, Pilani
K.K.Birla Goa Campus
Zuarinagar, Goa-403726, India
s_k_sahay@yahoo.com

## ABSTRACT
The dynamic web has increased exponentially over the past few years with more than thousands of documents related to a subject available to the user now. Most of the web documents are unstructured and not in an organized manner and hence user facing more difficult to find relevant documents. A more useful and efficient mechanism is combining clustering with ranking, where clustering can group the similar documents in one place and ranking can be applied to each cluster for viewing the top documents at the beginning.. Besides the particular clustering algorithm, the different term weighting functions applied to the selected features to represent web document is a main aspect in clustering task. Keeping this approach in mind, here we proposed a new mechanism called *Tf-Idf based Apriori* for clustering the web documents. We then rank the documents in each cluster using Tf-Idf and similarity factor of documents based on the user query. This approach will helps the user to get all his relevant documents in one place and can restrict his search to some top documents of his choice. For experimental purpose, we have taken the Classic3 and Classic4 datasets of Cornell University having more than 10,000 documents and use gensim toolkit to carry out our work. We have compared our approach with traditional apriori algorithm and found that our approach is giving better results for higher minimum support. Our ranking mechanism is also giving a good F-measure of 78%.

## 1. INTRODUCTION
World Wide Web is one of the most popular information resources for text, audio, video and metadata. The amount of data on the web has expanded many thousand times since its inception [1]. The modern search engines are faced with the enormous task of returning the few most relevant search results based on user query. In general the search results returned using any searching paradigm are not clustered automatically. But as the documents returned for a keyword may be of different nature depending upon the different meanings of the keyword. That is to say that the set of documents returned for a given keyword may further be subdivided into subsets of documents conveying similar sense of the keyword. Clustering the set of results will do this further sub-division and will present the results in a better way. It organizes the documents in such a way that the documents belonging to a group (cluster) are more similar to each other than to the ones which are a part of a different subgroup. Users often need to find the results related to a keyword pertaining to a particular meaning of that keyword. Since the documents which convey same meaning of the keyword will have similar words in them, they would automatically be grouped in the same cluster in most of the cases. Many mechanisms such as Decision trees, inductive logic programming, neural networks, rule-based systems, association techniques of data mining, genetic algorithms etc. are heavily used for web document clustering. All these techniques are most widely used for research areas such as information retrieval, database, machine learning, artificial intelligence and natural language processing. Many websites enable users to tag any web page with short free-form text strings, collecting thousands of keywords per day. Appropriate mining strategies, e.g., clustering are required for analysis of such tag information and its use in increasing the efficiency of the search engine. The clustering process is sometimes also called the unsupervised learning process because the class to cluster is not known at the time of creation of the cluster. Clustering helps to partition the input space into $k$ regions $C_1, C_2,…,C_k$ on the basis of some similarity metrics, where the value of $k$ may or may not be known previously. Several clustering algorithms are proposed in the literature [2]. These algorithms are divided into different types according to their nature of operation (e.g. Hierarchical, Partitional, Density-based, Grid-based, Graph-based, Prototype-based etc.).The web information usually is acceded by search engines and by thematic web directories. Search engines, such as Google, return to us a sorted list which besides the list of relevant documents they show us a cluster hierarchy. When thematic web directories are used, the documents are showed classified in taxonomies and the search process uses that taxonomy. In this context, the document clustering algorithms are very useful to apply to tasks such as automatic grouping before and after the search, search by similarity, and search results visualization on a structured way. Two aspects are very important in order to obtain good web page clustering results: the clustering algorithm, and the term weighting function applied to the selected features of the web pages. Ranking the documents inside each cluster further narrow down the user search. Many of the ranking algorithms are either content or linked based. Fig.1 shows the complete system architecture of clustering and ranking of web documents for any user query. In our approach we have proposed a technique called *Tf-Idf based Apriori*, which uses the threshold with the combination of Tf-Idf to make sets of frequent itemset on documents. The apriori algorithm [14] generally used for finding frequent itemsets in a database using candidate generation. Our frequent candidate itemset generation concept is same as frequent itemset generation and candidate itemset generation of traditional apriori algorithm. We are formulating the threshold as follows:

$$threshold = (1/\text{minimum support}) * \log_{10}(\text{total number of documents}/\text{minimum support}). \quad (1)$$

We use the above threshold to eliminate rows and columns of tf-idf table created during each frequent candidate itemset generation. For ranking the documents in each cluster, we applied the *cosine similarity*(discussed in sec 3.1.2) between every pair of documents in each cluster. Using this, we calculate the *similarity factor* of each document which shows how far a document similar to other documents in the same cluster and finally ranking has been done based on the user query. This can helps the user to find all his documents in an organized and ranked manner.

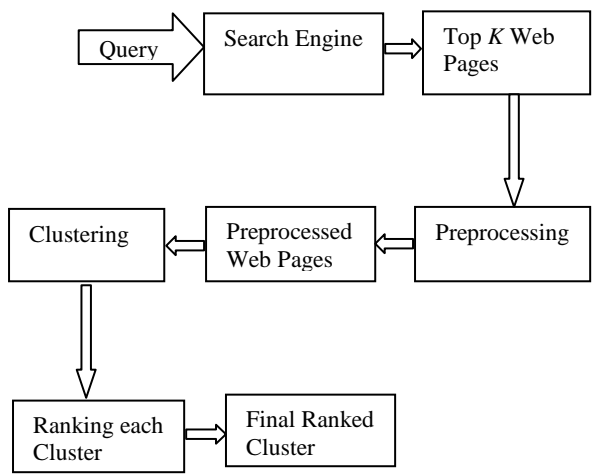

**Fig 1: System Architecture**

The remainder of the report is organized as follows: Section 2 covers the related work based on different clustering techniques used for web document. Section 3 describes the background details used in the proposed approach. In section 4, we describe the proposed approach adopted to form the clusters and ranking each cluster. Experimental work carried out by section 5 and finally section 6 describes the conclusion and future work.

## 2. RELATED WORK

Clustering and ranking are the two boosting and famous mechanism for extracting useful information on the web. In clustering an unstructured set of objects form a group, based on the similarity among each other. Among all the clustering algorithms one of the most likely algorithm is k-means. But the wrong choice of clusters(k) may produce wrong results, which is one of the problem of this algorithm. In case of an ambiguous query, word sense discovery is one of the useful methods for Information Retrieval, in which documents are clustered into a corpus. Discovering word senses by clustering the words according to their distributional similarity is done by Patrick et al. [3]. The main drawback of this approach is that they require large training data to make proper cluster and its performance is based on cluster centroid, which changes whenever a new web page is added to it. Hence identifying relevant cluster will be a tedious work. In 2008, Jiyang Chen et al. [4] purposed an unsupervised approach to cluster results by word sense communities. Clusters are made based on dependency based keywords which are extracted for large corpus and manual label are assigned to each cluster. To improve the cluster accuracy, Doreswamy et al.[5] developed a novel distance matrix which can integrated with k-means to give better clusters. Chakrabarti [6] also discusses various types of clustering methods and categorizes them into partitioning, geometric embedding, and probabilistic approaches. Data clustering is an established field. Mansaf Alam et al.[7] used heuristic search and LSI to cluster the web documents. Peng Li et al.[8] improved the web document clustering by using user related tag expansion techniques. Ingyu Lee et al.[9] proposed and approach for web document clustering based on bisection and merge. For efficient clusters their approach performs both bisection and merges steps based on normalized cuts on a similarity graph. R. Thiyagarajan et al. [10] proposed a new web recommended system using weighted k-means clustering algorithm which predicts the user's navigational behavior. Efficient phrased-based indexing has been used by Khaled et al. [15] for web document indexing. In this paper they discuss a novel phrase-based document model which when combines with an incremental document clustering algorithm based on maximizing the tightness of clusters, gives an improved results in web document clustering. B.Shanmugapriya et al.[16] describe an approach for effective distance measure using modified projected k-means clustering algorithm. Ranking the web documents also play a vital role in search processing. Many ranking algorithms [1, 18] have already been proposed like HITS(Hyper Induced Topic Search), WPR(Weighted Page Rank), WLR(Weighted Link Rank), WPCR(Weighted Page Content Rank), SALSA(Stochastic Approach for Link-Structure Analysis), Time Rank, Tag Rank etc.

Tf-Idf based frequent candidate itemset generation has been used in the proposed approach whose aim is to eliminate those itemset whose values are more than the pre-calculated threshold value. This process will continue till one not able to generate any further frequent candidate item sets. Finally one can get the clusters with similar documents. Then the ranking mechanism will applied on each cluster for better results. We use Gensim, a python toolkit to avoid the dependencies of the large training corpus size and its ease of implementing vector space model. The proposed approach has been compared with the traditional apriori algorithm. Results show that our approach can outperform the traditional apriori algorithm even when the minimum support is high. The ranking mechanism which has used for each cluster to the rank the documents also giving a better performance in terms of F-measure.

## 3. BACKGROUND
### 3.1 Vector Space Model

Vector space model(VSM) [11] is a popular, most widely used algebraic model for representing text documents as vector of identifiers. Here, every document can be represent as a multidimensional vectors of keywords(i.e keywords extracted from that document) in Euclidean space. The weight associated with each keyword determines the relevance of the keyword in the document. Hence a document in vector form can be represent as, $D_j = [w_{1j}, w_{2j}, w_{3j}, w_{4j}..., w_{nj}]$ where, $w_{ij}$ is the weight of keyword i in document j.

*3.1.1 TF-IDF*

TF-IDF is generally a content descriptive mechanism for the documents. The term frequency (TF) is the number of times a term appears in a document and is calculated as follows: tf = (Number of occurrences of the keyword in that particular document) / (Total number of keywords in the document). Inverse Document Frequency(IDF) measures the rarity of a term in the whole corpus. Denoting the total number of documents in a collection by N, the inverse document frequency of a term t is defined as follows: idf = log ( N / df ). The concepts of term frequency and inverse document frequency [11] are combined, to produce a composite weight for each term in each document. tf-idf = tf * idf.

*3.1.2 Cosine-Similarity Measure*

There are many techniques to measure the similarity between the user query and the retrieved documents. One of such widely used technique is cosine-similarity [11]. It is one of the powerful similarity checking technique compare to all the other techniques exist[14] and widely used for web document similarity. Cosine-similarity(q,d) = $\frac{q.d}{||q||*||d||}$    (2) where, q and d are query and document vectors respectively. Also ||q|| and ||d|| represent their length respectively. The strength of the similarity depends on the value of θ. If $θ = 0^0$, then the document and query vector are similar. As θ changes from $0^0$ to $90^0$, the similarity between the document and query decreases.

### 3.2 Gensim
Gensim[12] is a python library and mainly use for vector space modeling. It's basic use is for Natural Language Processing(NLP) community and can process raw, unstructured digital text. Because of its memory independent features, it can handle large web based corpora and also many vector space algorithm. It can automatically extract semantic topics from web documents and having many other salient features.

### 3.3 The Apriori algorithm
**Input:** The dataset (D) and min_sup.
**Output**: The frequent itemset.
1. k = 1;
2. Find frequent itemset, $L_k$ from $C_k$, the set of all candidate itemsets;
3. Form $C_{k+1}$ from $L_k$;
4. k = k+1;
5. Repeat 2-4 until $C_k$ is empty;

Step 2 is called the frequent itemset generation step. Step 3 is called as the candidate itemset generation step. Details of these two steps are described below.

**Frequent itemset generation**
  Scan D and count each itemset in $C_k$, if the count is greater than min_sup, then add that itemset to $L_k$.

**Candidate itemset generation**
  For k = 1, $C_1$ = all itemsets of length = 1.
  For k > 1, generate $C_k$ from $L_{k-1}$ as follows:
The join step:
    $C_k$ = k-2 way join of $L_{k-1}$ with itself.
    If both $\{a_1,...,a_{k-2}, a_{k-1}\}$ & $\{a_1,..., a_{k-2}, a_k\}$ are in $L_{k-1}$,
    then add $\{a_1,...,a_{k-2}, a_{k-1}, a_k\}$ to $C_k$.
    The items are always stored in the sorted order.
The prune step:
    If any non-frequent (k-1) subset found in $\{a_1, a_2, a3...........a_k\}$ then discard this set.

## 4. PROPOSED APPROACH
### 4.1 Cluster Formation:

**Input:** minimum support and the user query.
**Output:** Number of Clusters each having documents in ranked form.

1. Web page extraction and preprocessing: Submit the query to a search engine and extract top 'N' pages. Preprocess the retrieve corpus as follows:
   - Remove the stop and unwanted words.
   - Select noun as the keywords from the corpus and ignore other categories, such as verbs, adjectives, adverbs and pronounce.
   - Do stemming using porter algorithm [13].
   - Save each preprocessed 'N' pages as documents $D_i$, where i = 1, 2, 3,. . . ,N.

2. After keyword extractions, consider each keyword as a transaction and the documents $D_i$ in which the keyword occurs as transaction elements.

3. Calculate *tf* for each distinct keyword in each $D_i$ as,
     *tf* = 1/ (Number of distinct keywords in a document)
   Calculate *idf* for each distinct keyword in each $D_i$ as,
     *idf* = $\log_{10}$ (total number of documents/number of documents the keyword appears in)

4. Calculate *tf*idf* value for each distinct keyword in each $D_i$ and represent all the values in the *tf*idf* table.

5. Calculate threshold as, threshold = (1/minimum support)*$\log_{10}$(total number of documents/minimum support)

6. Generate *n* frequent candidate itemsets (S, where n >= 2) for keywords till, 0<min {$tf*idf_{D1}, tf*idf_{D2},\ldots, tf*idf_{DN}$} <= threshold for all generated S and at each step do the followings:

   Calculate *tf* as, *tf* = 1/ (number of times S appears in the document)for each *n* frequent candidate itemset in each document.

   Calculate *idf* as, *idf* = $\log_{10}$ (total number of documents/number of documents S appears in) for each *n* frequent candidate itemset.

   Calculate *tf*idf* value for each *n* frequent candidate itemset in each document and represent all the values in the *tf*idf* table.

   Now mark the '*n*' frequent candidate itemsets (rows) for elimination if min { $tf*idf_{D1}, tf*idf_{D2},\ldots, tf*idf_{DN}$ } > threshold.

   mark documents (columns) for elimination if min { $tf*idf$ n frequent candidate item set1 , $tf*idf$ n frequent candidate item set2 , · · · . . . , $tf*idf$ n frequent candidate item setN} >threshold.

7. Final Clusters ($C_i$) where i = 1, 2, 3,. . . ,M formed each having group of similar documents($D_k$) where k=1…N and N may vary from cluster to cluster.

## 4.2 Ranking Clusters

**Input:** Final Clusters ($C_i$) where i = 1, 2, 3,. . . ,M formed by above approach and the user query $W_e$ .

**Output:** Ranked Clusters.

1. Preprocessed user query words $W_e$, where e = 1, 2, . . .,p.

2. i) Compute the cosine similarity between every pair of document($C_i$) *i. e* $\forall$i,j cosinesim($D_i, D_j$), 1≤i≤N and 1≤j≤N using Eq. 2.

  ii) Compute similarity factor of each $D_k$ as follows:
*Simfact($D_k$)* =
$\sum_{m=1, m \neq k}^{N} [\left(\frac{No.of\ keywords\ common\ to\ Dk\ and\ Dm}{Total\ unique\ words\ in\ Dk\ and\ Dm}\right) * cosinesim(D_k, D_m)]$

3. Ranking of each document is as follows:
   Rank($D_k$) = $(\sum_{e=1}^{p}(Tf*Idf)W_e, D_k) *$ Simfact($D_k$), where (Tf*Idf)$W_e$, $D_k$ represents (Tf*Idf) of Keyword $W_e$ with respect to the document $D_k$.

4. Sorting the documents in each $C_i$ based on their ranked values will give the desire output.

## 5. EXPERIMENTAL RESULT

For experimental purpose, to demonstrate the effectiveness of our approach, we have taken the Classic3 and Classic4 datasets of Cornell University [17]. We consider the top 400 documents and tested both Tf-Idf apriori approach and traditional apriori approach on it. The results are shown in Fig. 2. We found that when the minimum support increases our approach still detect more clusters than traditional apriori algorithm. We have tested the F-measure [19] of our ranking mechanism of documents inside each cluster. It measures the system performance by combining Precision and Recall. It represents the harmonic mean of Precision and Recall. In F-measure, Precision and Recall are evenly weighted thus reflecting overall performance of the algorithm under consideration.

$$\text{F-measure} = \frac{2 \times Precision \times Recall}{Precision + Recall}$$

For demonstration purpose we have shown F-measure of some of the clusters in Fig. 3. The results show that on an average 78% of documents have ranked in a proper order in each of the cluster.

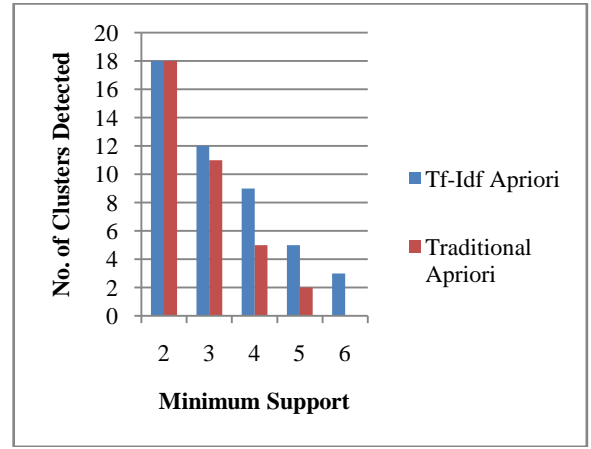

**Fig 2: Tf-Idf Apriori v/s Traditional Apriori**

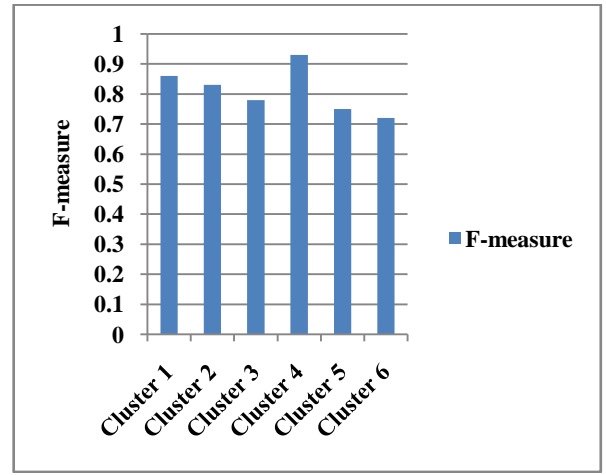

**Fig 3: F-measure of different clusters**

## 6. CONCLUSION AND FUTURE WORK

In this paper, we proposed an approach called Tf-Idf based Apriori. An equation has been formulated for finding the threshold which when combine with our modified Tf-Idf can able to identify frequent itemsets on a set of documents. We use this threshold to eliminate rows and columns of tf-idf table created during each frequent candidate itemset generation. This frequent candidate itemset generation concept which has been used in our approach is same as frequent itemset generation and candidate itemset generation of traditional apriori algorithm. In experimental work we have consider the Classic3 and Classic4 datasets of Cornell University and taken the top 400 documents to compare our approach with traditional apriori algorithm and found that our approach is giving better results than traditional apriori algorithm. For ranking the documents in each cluster, we applied the cosine similarity between every pair of documents in each cluster. Using this, we calculate the similarity factor of each document and finally rank their values. We found that on an average 78% of documents have ranked in a proper order in each cluster. This ranking of documents will help the user to get the necessary documents at the beginning of each cluster and reduced his search process. This work can further be extended by considering those documents which are not parts of the initial clusters formed by the proposed approach because of strong association rule, to make either new clusters or part of the existing clusters which may be of user interest.

## 7. ACKNOWLEDGEMENTS

We sincerely thank to our colleagues Bharat Deshpande, Biju K. Ravindran and Neena Goveas for their useful discussions and valuable suggestions.

## 8. REFERENCES

[1] Naresh Barsagade. Web usage mining and pattern discovery: A survey paper. CSE8331, Dec, 2003.

[2] A. Jain, M. N. Murty, and P. Flyn. Data clustering: A review. ACM Computing Surveys, vol.no. 3, pp. 264323, 1999.

[3] Patrick Pantel and D. Lin. Discovering Word Senses from Text, SIGKDD'02, July 23-26, 2002, Edmonton, Ablerta, Canada.

[4] J. Chen, O. R. Zaiane and R. Goebel. An Unsupervised Approach to Cluster Web Search Results based on Word Sense Communities, 2008 IEEE/WIC/ACM, International Conference on Web Intelligence and Intelligent Agent Technology.

[5] Doreswamy and Hemanth K.S. A Novel Design Specification(DSD) based K-mean clustering performance Evaluation on Engineering material's database, IJCA, Vol 55, No.15, Oct-2012.

[6] S.Chakrabarti. Mining the Web: Discovering Knowledge from Hypertext Data. Morgan Kaufmann Publishers 2003.

[7] Mansaf Alam and Kishwar Sadf. Web search result clustering using heuristic search and latent semantic indexing, IJCA, Vol 44, No.15, April-2012.

[8] Li p, Wang B and Jin W. Improving web document clustering through employing user-related tag expansion techniques.Journal of Computer Science and Technology 27(3):554-556 May-2012. DOI 10.1007/ S11390-012-1243-4.

[9] Ingyu Lee and B-Won. An Effective web document clustering algorithms based on bisection and merge, Artif Intell Rev(2011) 36-69-85, DOI 10.1007/S10462-011-9203-4.

[10] R.Thiyagarajan, K.Thangavel and R.Rathipriya. IJCA Vol-86, No.14, Jan-2014.

[11] http://www.miislita.com/term-vector/term-vector-3.html

[12] http://www.nlp.fi.muni.cz/projekty/gensim/intro.html

[13] http://tartarus.org/martin/PorterStemmer/def.txt

[14] Introduction to Data Mining by P.Tan, M. Steinbach and Vipin Kumar, Pearson Education, 2006.

[15] Khaled M and Mohamed S. Efficient phrased-based document indexing for web document clustering, IEEE Transaction on Knowledge and Data Engineering, Vol 16, No. 10, October-2004.

[16] B. Shanmugapriya and M.Punithavalli. A modified projected K-means clustering algorithm with effective distance measure, International Journal of Computer Application, Vol 44, No.8, April-2012.

[17] http://www.dataminingresearch.com/index.php/2010/09/classic3-classic4-datasets

[18] Jaroslar Pokorhy, Jozef Smizansky. Page Content Rank, An approach to the web content mining.

[19] http://en.wikipedia.org/wiki/F1_score